\documentstyle[12pt,epsf]{article}

\newcommand{\be}{\begin{equation}}
\newcommand{\ee}{\end{equation}}

 1
\font\elevenrm=cmr10 scaled\magstep 1
 1

\def\ref{\hang\noindent}

\begin{document}
\vspace*{1.8cm}
  \centerline{\bf SOME EARLY RESULTS FROM THE}
  \centerline{\bf ROSSI X-RAY TIMING EXPLORER (RXTE)}
\vspace{1cm}
  \centerline{HALE BRADT}
\vspace{1.4cm}
  \centerline{MASSACHUSETTS INSTITUTE OF TECHNOLOGY}
  \centerline{\elevenrm Room 37-587, Cambridge MA 02139-4307, USA}
\vspace{3cm}
\begin{abstract}
The \it Rossi X-ray Timing Explorer (RXTE) \rm was successfully launched on 1995 December 30 and has been operational since that time. Its three instruments are probing regions close to compact objects, degenerate dwarfs, neutron stars, stellar black holes and the central engines of AGN. Temporal studies with the ASM and PCA have already yielded rich results which pertain to the environs, evolution, and nature of the compact objects in galactic systems. Here I review some selected results from these instruments. as obtained by various RXTE observers. The ASM is providing detailed light curves of about 60 detected sources and has uncovered new temporal/spectral states of galactic binary systems. The bizarre behavior of the possibly very young binary system, Cir X-1, is being revealed in detail by both the ASM and the PCA. A rare high/soft state of the black-hole candidate, Cyg X-1 provides new insight into the nature of the low/high transitions in black-hole binaries. The PCA has made possible the discovery of oscillations near 1 kHz in ten low-mass X-ray binary systems. These are most probably direct indicators of the neutron-star spin in some cases and probably indirect indicators in others. Studies of microquasars have unveiled a host of new temporal phenomena which may provide links between accretion processes and the radio jets in these systems.
\end{abstract}
\vspace{2cm}
\section{Objectives, Instruments, and Launch}

The scientific objective of the \it Rossi X-ray Timing Explorer (RXTE) \rm is to study the nature, environs, and evolution of galactic and extragalactic compact objects/systems through the study of the temporal variation of the emerging radiation over a very wide band of X-ray energies (2--200 keV). The compact objects being studied include galactic and extragalactic objects, namely white dwarfs, neutron stars, stellar black-hole candidates, and active galactic nuclei (massive black holes).\footnote{This paper was completed late in 1996; some results herein reflect work postdating the conference. This paper also will appear in the Proceedings of 5th International Workshop on Data Analysis in Astronomy, CCSEM Center, Erice, Italy, (Oct. 1996), eds. L. Scarsi and C. Maccarone, World Scientific Publ. Co., with the much appreciated permission of the Editors of both proceedings.}

The three instruments on \it RXTE \rm and the spacecraft capabilities are described by Bradt, Rothschild \& Swank (1993). Two pointed instruments feature, respectively, proportional counters and crystal scintillators with large effective areas, 0.70 m$^2$ and 0.16 m$^2$, respectively. Their angular resolution (1$^\circ$ FWHM, circular) is obtained with mechanical collimation. The large areas and rapid onboard data processing permit  the measurement of large count rates up to high energies. The high statistics obtained from bright sources permits the study of intensity variations on time scales as short as a few microseconds. 

The proportional counter array (PCA) system consists of 5 large detectors with sensitivity over 2--60 keV. The crystal system (HEXTE) consists of two clusters of four phoswiches each with a nominal response of 15--200 keV. The latter clusters rock on or off the target every $\sim$15-s to obtain frequent background measures with phasing such that one cluster is always viewing the target.

The third instrument is an All-Sky Monitor (ASM). It surveys up to about 80\% of the sky with 5--10 samples of any celestial region each day. The overall energy range of the ASM is 1.5--12 keV which is telemetered in three energy channels. It can monitor sources in uncrowded regions down to about 35 mCrab (2 $\sigma$) in one 90-s exposure or about 10 mCrab for 1-day averages. The data and light curves from the ASM are public and available on the web: 

http://space.mit.edu/XTE/XTE.html 

http://heasarc.gsfc.nasa.gov/docs/xte/asm\_products.html.

The spacecraft can point the PCA/HEXTE to any position on the sky anytime of the year except that directions within $30^\circ$ of the sun are excluded. A new target can be acquired within hours of a celestial event detected with the ASM or by other ground-based or satellite-borne instruments. This capability has been used to great advantage on \it RXTE, \rm though usually with a response time on the order of 24 hours because a more rapid response was not required.

The \it RXTE \rm was launched successfully on 1995 December 30. All three instruments are currently operational with only relatively minor deficiencies; they are performing very close to their full design capabilities.

In the following, I present some early results that are forthcoming from \it RXTE \rm. I do not attempt to present a balanced overview of the \it RXTE \rm output; the studies being carried out by hundreds of observers are too wide ranging and diverse. Rather, I select certain topics in which I have been involved or which are of particular interest to me. The results given here are mostly temporal studies from the ASM and the PCA; these results are more easily forthcoming than spectral studies (including HEXTE data) because the high quality calibrations required for spectral analysis are only now becoming available. The Guest Observing Program started in early in 1996 February and AO-2 observations began in 1996 November. Early results from \it RXTE \rm appeared in the 1996 Sept. 20 issue of Astrophysical Journal Letters.

\section{Light Curves from the All-Sky Monitor}

Two of the three ASM detectors developed high-voltage breakdown shortly after launch, and all three detectors were shut down for diagnostics. The working counter was reactivated on 1996 Feb. 22, and essentially complete light curves together with 3-channel hardness ratios were obtained from that date. The two failed detectors were recovered with most of their effective area by mid-March 1996.

Early results from the ASM have been reported by Levine et al. (1996). From a catalog of $\sim$170 cataloged sources, about 60 yield day-by-day light curves and an additional $\sim$40 are possibly detected when averaged over a number of days. The light curves show a wide variety of behavior. Six examples of light curves of galactic sources for the year 1996 are shown in Fig. 1. All are accreting binaries. The compact object is believed to be a neutron star for the three systems portrayed on the left side of the figure and a black hole for the three on the right. The intensity is given in ASM ct/s in which the Crab nebula yields 75 ct/s. Features of note for these sources are described below. See Levine et al. (1996) and van Paradijs (1995) for references to previous work on these sources.

Cir X-1: Circinus X-1 is believed to be an LMXB in a 16.6-d elliptical orbit as evidenced by periodic outbursts that are presumed to occur during periastron passage. (See refs. in  Shirey et al. 1996.) These ASM data show it to be continuously active during 1996 with the 16.6-d outbursts highly visible. (It can be very faint or undetectable for many months at a time.) The nature of this system with its highly obscured IR counterpart is not well known (see, e.g., Glass 1994). It may be a young runaway system from a nearby supernova remnant (Stewart et al. 1993). This source is discussed further below. 

SMC X-1: The periodic on-off behavior of this source confirms the $\sim$60-d cyclic period suggested by previous observers (Gruber \& Rothschild 1984). If so, this is likely analogous to the 35 d cycle of Her X-1 and the 30-d cycle of LMC X-4 which are attributed to precession of the accretion disk (see Priedhorsky \& Holt 1987).

Cyg X-2: This evolving double-humped light curve has an apparent 78-d period (Wijnands, Kuulkers, \& Smale 1996a). This period is independently seen in Ariel 5 and Vela 5B data (e.g., Smale \& Lochner 1992). This could be another precessing disk, but it does not follow the behavior observed in other known precessing-disk systems.

Cyg X-1: This black-hole candidate (BHC) entered a rarely observed high (soft) state in May 1996 with a peak intensity of $\sim$1.3 Crab. It returned to its low state in August. The spectral change was dramatically evident in the ASM hardness ratios and also in a comparison with BATSE data. The entire spectral range of the latter is above the ASM range. In both instruments, the high-energy flux was found to decrease. (Cui et al. 1997a, Zhang et al. 1997). There is more on this source below.

GRS 1915+105: This microquasar shows a remarkable series of intensity/spectral states; it is discussed below.

4U 1630--472. This soft X-ray transient may contain a black hole because its X-ray spectrum is very soft with a hard tail reminiscent of Cyg X-1 (Parmar, Stella, \& White 1986). These data show the source appear and then disappear again about 150 days later. The turn on in such systems may be  triggered by an accretion-disk instability (see, e.g., Mineshige \& Wheeler 1989). These data show the turn-on and turn-off in previously unseen detail, which should be useful in distinguishing models.

Reports of periodic behavior from the ASM data include the confirmations of the $\sim$78-d period in Cyg X-2 and the $\sim$60-d period in SMC X-1 noted above, a new $\sim$37-d period  in Sco XÐ1 (Peele \& White 1996), confirmation of a 2.7-h period in 2S 0114+650 together with an 11.7-d modulation consistent with the optical period (Corbet \& Finley 1996), a 5.6-d orbital period in X-rays in Cyg X-1 (Zhang, Robinson, \& Cui 1996a) also reported in Ginga data, and a 24.7-d modulation in the low-mass X-ray binary GX 13+1 also apparently seen in Ariel 5 data (Corbet 1996). Most of these reports have not yet reached the refereed literature and hence some should be viewed with caution. 

The ASM has not yet discovered a completely `new' X-ray source. Nevertheless it is proving invaluable (1) in illuminating the character of the long-term variations of various types of sources, (2) in discovering new or unusual states of sources, (3) in guiding the observing plan of the observatory, and most important, (4) in providing a temporal/spectral context for the relatively brief observations carried out by the PCA/HEXTE of a given source.

\section{Cir X-1: QPOs and Rapid Absorption Events}

As noted above, Circinus X-1 has been continuously active since the launch of \it RXTE.\rm. An expanded ASM light curve and plots of hardness ratios are shown for six cycles of its 16.6-d period in Fig. 2. Repeatable systematic variations of the hardness ratios are readily apparent (Shirey et al. 1996). This variation in hardness ratios had been reported previously but only as an average over many cycles in Ginga data (Tsunemi et al. 1989).

Circinus X-1 has long been known to exhibit rapid variability on time scales down to less than 1 s (see e.g., Dower, Bradt, \& Morgan 1982, Oosterbroek et al. 1995 and refs.  therein). Observations with the high statistics obtainable with the \it RXTE \rm PCA add substantial insight into the character of this variability. Three segments of \it RXTE \rm data are shown in Fig. 3; dramatic dips are apparent in the two top segments. The decreased flux during a low state (presumably during a dip) and the recovery from it have been shown by Brandt et al. (1996) with ASCA data to be due to changing absorption with partial covering. The \it RXTE \rm data confirm this behavior over a wider range of intensity/temporal conditions and X-ray bandwidth. The third segment of Fig. 3 shows a steady flux but with variations well in excess of the $<$1\% expected from the statistics (13,000 ct/bin), indicating the presence of rapid non-statistical fluctuations.

Power density spectra (PDS) of this quiescent flux (Fig. 4) show substantial power in excess of Poisson fluctuations (Shirey et al. 1996). The non-Poisson variations include pronounced quasi-periodic oscillations (QPO) of relatively high Q which drift between 1.3--12 Hz, a flat-topped noise, and a broad peak (QPO) with centroid that varies from 20--100 Hz in the PDS of Fig. 4. 

The evolution of source characteristics (frequency and energy spectra) during the quiescent periods between outbursts (Fig. 4) are also reported by Shirey et al. (1996). The features of the PDS (e.g., the frequencies of the two QPO and the level of the flat-topped power) are remarkably correlated with one another. These characteristics also appear to be correlated with phase of the 16-d orbit, but with some notable deviations. Some of these temporal features were previously known (e.g., Tennant 1988, Oosterbroek et al. 1995), but the high sensitivity and repeated observations of \it RXTE \rm permit one to follow their evolution with orbital phase. The conjecture is that the evolution during this quiescent phase reflects accretion from the disk while it is not being replenished from the normal companion.

\section{Cygnus X-1: Transitions to and from the High State}

Spectral and timing studies of Cyg X-1 were carried out periodically with the \it RXTE \rm/ PCA during the transition to the high/soft state. The data are public and two groups studied them (Belloni et al. 1996; Cui et al. 1997a). Their results provide insight into the spectral-formation processes. Both groups report that the spectral softening in the transition was associated with an increase in the high-frequency cutoff in the PDS of the X-ray signal, not inconsistent with the behavior of other black-hole candidates. The latter authors note that this is consistent qualitatively with a Comptonizing corona which reduces size (but not density) as the source moves into the high state. A smaller corona means fewer inverse-Compton scatters. This results in relatively fewer high energy photons, i.e., a softening spectrum. It also leads to less smearing of the high-frequency fluctuations in the source photons; i.e., the cutoff frequency in the PDS increases.

This scenario was reinforced dramatically with studies of the cross-correla-tion between hard and soft fluxes as the source moved into and out of the high state (Cui et al. 1997b). A delay of hard photons relative to soft photons had been reported by Miyamoto et al. (1988) for Cyg X-1. The \it RXTE \rm data exhibited this delay and further showed that the average delay decreased during the transition to the high state, becomes very small during it, and increased again during the return to the low state (Fig. 5). In general, hard-flux delays are expected from the Comptonization up-scatter process, and one would expect these delays to be reduced for a smaller corona with fewer scatters, in agreement with the data. 

\section{Variable QPO Oscillations at $\sim$1 kHz (Sco X-1)}

The fast spin of millisecond (1--10 ms) radio pulsars has long been postulated to be due to gradual angular momentum transfer by accreting matter in low-mass X-ray binaries (Smarr \& Blandford 1976; Bhattacharya 1995). However, such pulsations had not been detected from X-ray binaries before the launch of \it RXTE \rm. A prime objective of \it RXTE \rm was to find this missing link in the evolution of neutron-star systems. It now appears that \it RXTE \rm has been successful in this endeavor. There are apparently two types of such oscillations: variable-frequency quasi-periodic (QPO) oscillations in the persistent flux and coherent oscillations during X-ray bursts. Both may be ramifications of the neutron-star spin.

At this writing, there are eight X-ray sources for which variable-frequency QPOs have been reported: 4U 1728--34 (Strohmayer et al. 1996), Sco X-1 (van der Klis et al. 1996a), 4U 1636--536 (Zhang et al. 1996b), 4U 1608--52 (Berger et al. 1996), 4U 0614+091 (Ford et al 1997), 4U 1735--44 (Wijnands et al. 1996b), 4U 1820--30 (Smale, Zhang, \& White 1996), and GX 5--1 (van der Klis et al. 1996b). Here I will describe the oscillations from Sco X-1.

The brightest X-ray source in the sky is Sco X-1; hence it provides the high statistics in short periods needed in an efficient search for, and tracking of, high-frequency QPO. A double QPO feature has been found in this source (Fig. 6; van der Klis et al. 1996a); with peaks at $\sim$800 and $\sim$1100 Hz. The peaks are not always present and sometimes only the 1100-Hz peak is present. The frequencies of the two peaks increase with accretion rate. The high statistics allow the frequencies of the QPO to be tracked with time. With longer integration times, the peaks would be substantially broadened or washed out. The power in these peaks is not large; the fractional rms amplitudes are of order 1\%.

The frequency variation of these QPOs precludes interpreting the period directly as the neutron-star spin frequency. Nevertheless the high frequencies suggest strongly that the rapid oscillations originate in the immediate region of the neutron star. One possibility proposed by van der Klis et al. (1996a) is that the 1100 Hz represents the Kepler frequency of matter in the inner disk and that the 800 Hz is the beat frequency between the Kepler frequency and the neutron-star spin. If so, the spin frequency is $\sim$250 Hz. Several of the other sources listed above also exhibit two QPO peaks that move in frequency as the intensity varies. It appears that a number of these sources are exhibiting the same phenomenon.

A rather explicit model for the production of kilohertz oscillations in a neutron-star system has been put forward by Miller, Lamb \& Psaltis (1997). They propose that the detected oscillations are at the Kepler frequency of the marginally stable orbit, from which clumps of matter are gradually stripped to create a stream of matter falling onto the neutron star. The initial radius of the clump is the \it sonic radius \rm. The result is a hot footprint on the neutron star which rotates around the star (in the inertial frame) at the Kepler frequency of the sonic radius. This frequency is thus directly measurable. In this model, the detection of a high Keplerian frequency places an upper bound on both the mass and radius of the neutron star. In turn, this places constraints on the equation of state in all neutron stars.

Other proposed origins of kilohz oscillations are photon-bubble oscillations in the accretion column (Klein et al. 1996) and neutron-star vibrations (e.g., McDermott, Van Horn, \& Hansen 1988). The latter origin seems excluded because it does not naturally yield the substantial dependence of the QPO frequencies upon mass accretion rate.

\section{Oscillations during Bursts at $\sim$1 kHz (4U1728--34)}

In work that paralleled that just described, Strohmayer et al. (1996) discovered oscillations during an X-ray burst from the LMXB source 4U1728--34. This source is a well known Type I X-ray burst source; i.e., its atmosphere occasionally undergoes unstable thermonuclear He burning. Eight bursts were detected during the Strohmayer et al. (1996) observations, and six of them showed oscillations near 363 Hz. 

The frequency evolution of the 363-Hz signal in one burst is shown in Fig. 7 together with the light curve of the burst itself. The frequency varied $\sim$1.5 Hz as the burst progressed becoming effectively coherent in the burst tail at 364 Hz. The pulsations were present during the rise of the burst; fell below threshold near the peak, and reappeared thereafter. The rms amplitude was $\sim$10\% at the start of the rise and 3--7\% after reappearing. The authors propose that these oscillations reflect directly the neutron-star spin period. The high coherence of the 364 Hz in the tail gives credence to this interpretation. 

The \it persistent \rm emission from this source (4U 1728--34) exhibits two high-frequency QPO in the PDS, similar to those of Sco X-1. The higher frequency varies from 700--1100 Hz and the lower, when present, varies between 600 and 800 Hz. The rms amplitude of these pulsations is in the 5--8\% range, a substantially larger portion of the flux than for Sco X-1. When both peaks are present, the difference of their frequencies is $\sim$360 $\pm$ 10 Hz, consistent with a constant frequency difference as the individual peaks vary in frequency. In the beat-frequency picture given above for Sco X-1, this would be the neutron-star spin frequency. Remarkably, it is in agreement with the frequency of the coherent frequency found during the burst. This strengthens the case for an underlying fundamental frequency of 364 Hz, and this most likely is the neutron-star spin frequency.  If so, a major goal of \it RXTE \rm has indeed been accomplished.

The $\sim$363-Hz modulation during the burst could arise from thermonuclear-flash inhomogeneities on the neutron-star surface (Bildsten 1995); the frequency drift during the burst rise could be due to the progression of the burning front on the surface (Strohmayer et al. 1996). 

We note that 4U0614+091 may be another case where the difference frequency between two high-frequency QPOs also appears directly, but in this case the latter oscillations are in the persistent flux at 328 Hz at only 3.3 $\sigma$ (Ford et al. 1997). Two other cases of oscillations in bursts have been found at this writing. The transient source KS 1731--30 exhibits 524 Hz oscillations for a duration of 2 s with Q $\geq$ 900 (Smith, Morgan, \& Bradt 1997; Fig. 8), and an unidentified source near GRO J1744--28 exhibited 589 Hz in a 4-s interval (Strohmayer, Lee, \& Jahoda 1996).

\section{Microquasars}

There are at least eight galactic X-ray sources that exhibit radio jets or their temporal/spectral characteristics, namely the long known Cyg X-3, SS433, and Cir X-1 (See refs. in van Paradijs 1995), the galactic-center sources 1E1740.7--2942 and GRS 1758--258 (see refs. in Mirabel \& Rodriguez 1994), the hard X-ray transient GRS 1739--278 recently discovered with Sigma/Granat (Vargas et al. 1997) and determined to be a variable radio flaring source (Hjellming et al. 1996), and finally two radio jet sources that exhibit superluminal motion: GRS 1915+105 (Mirabel \& Rodriguez 1994) and GRO 1655--40 (Tingay et al. 1995; Hjellming \& Rupen 1995). The latter two sources have been termed microquasars; in fact, all sources exhibiting radio jets may well deserve this name. These systems could well give insight into the relation between accretion processes and radio jets. The ASM data show that GRS 1915+105 has been active all year (see above) and that GRO 1655--40 became active on Apr. 25 of this year. Here I would like to present some remarkable \it RXTE \rm results on the two latter microquasars.

\subsection{GRS 1915+105}
This transient source, originally discovered in X rays with \it Granat \rm WATCH (Castro-Tirado et al. 1992) has been bright and active since the beginning of the \it RXTE \rm mission (Fig. 1). The ASM and PCA exhibit several distinct states, described and named in Morgan, Remillard, and Greiner (1997; Fig. 9). Observations with the high sensitivity of the PCA (Greiner, Morgan, \& Remillard 1996) show dramatic dips, with quasi-periodic repetitions on time scales of a few minutes (Fig. 10) during the `chaotic' state. These dips do not show the spectral or temporal characteristics expected of absorption events suggesting strongly that they are accretion phenomena. In the `bright' state, the source exhibits intense, relatively high-Q QPOs at 0.05--1 Hz (Morgan, Remillard, \& Greiner 1997; Fig. 11). These vary in frequency, disappear, and reappear at different frequencies. The stronger peaks have rms amplitudes of order 10--15\%.

The authors of the latter work have succeeded in tracking the phases of the individual cycles of some of these QPO. They find a random walk, in that the duration of one QPO cycle is uncorrelated with the duration of the adjacent cycles. Neither is it correlated with the amplitude of the cycle. These features eliminate classes of models such as those that invoke a reservoir or those predicting trains of relatively coherent (or gradually changing) pulses. Superposition of the individual pulses showed a significant delay in the hard pulsing, of $\sim$4\% of the pulse period, for periods of 1/2 to 15 s. They also showed full-width pulse amplitudes up to $\sim$40\% of the mean flux at the highest energies, $\sim$15 keV. The authors state ``that if the high-energy photons are derived from inverse Compton scattering, then [these characteristics] suggest that the energy distribution of the energetic electrons must be oscillating at the QPO frequency.'' Therein lies the possible connection to the radio jets.

The same work revealed a QPO at a higher frequency, 67 Hz, which reappeared six times at the same frequency (Fig. 12). It is quite a sharp feature (Q $\sim$ 20) with a relatively small rms amplitude, $\sim$1\%. It too is most powerful at high energies; on May 5, it had rms amplitude of 1.5\% below 4 keV and 6\% above 10 keV. The authors speculate that this frequency is that of the innermost stable orbit around a non-rotating black hole, which is at 3.0 Schwarzschild radii (3.0 $R_S$). In this case, the mass of the black hole is determined to be 33 $M_{\odot}$. Of course, if the black hole has significant angular momentum, this value would be different. An alternate explanation invokes transverse `diskoseismic' g-mode oscillations of the accretion disk (Nowak et al. 1997). However, this does not appear to be in accord with the very hard spectrum of the 67-Hz feature (R. Remillard, pvt. comm.).

Finally, we call attention to two other works on \it RXTE \rm data from this source during this time frame: Chen, Swank, \& Taam (1997) and Belloni et al. (1997). These bring somewhat different perceptions to the same phenomenology.

\subsection{GRO 1655--40}

The other microquasar, GRO J1655--40 is a transient source also discovered in X rays (Zhang et al. 1994), with CGRO/BATSE in 1994. It has exhibited subsequent outbursts, but was undetectable in the ASM in early 1996. However, it suddenly reappeared on 1996 Apr. 25 and reached $\sim$2 Crab in the next 10 days (Fig. 13). The compact object in this binary system has been found from optical studies of the companion to have a mass of 7.0 $\pm$ 0.2 $M_{\odot}$ (statistical errors only) which makes it a likely black hole (Orosz \& Bailyn 1997). 

The resurgence of this microquasar in April led to studies with the PCA. A high frequency QPO has been found at a frequency of $\sim$298 Hz when PDS are superimposed from the portions of the data exhibiting the hardest spectra (Remillard 1996). If, again, the periodicity is taken to be that of the innermost stable orbit of a non-rotating black hole, at 3.0 $R_S$, one obtains a black-hole mass of 7.4 $\pm$ 0.2 $M_{\odot}$ (Remillard 1996). This agrees, amazingly, with the optically measured mass of the black hole, 7.0 $\pm$ 0.2 $M_{\odot}$ quoted above. Turning the argument around, this would constitute a measurement of the angular momentum of the black hole. Unfortunately, since the scenario is not at all well established, this remains speculative, but the close agreement of the two mass determinations is quite provocative.

Just prior to the April resurgence, the source was being observed with optical photometry (Orosz et al. 1997). The multicolored ellipsoidal light curves showed a steady brightening in B, V, R, and I that began about six days before the X-ray flux began to rise. The onset began in the I band and then, sequentially, in the R, V, and B  bands over a period of 1.1 d. The authors suggest that this indicates an inward moving (`outside-in') disturbance in the accretion disk. They further suggest that the substantial delay before the X-ray onset might provide indirect support for advection-dominated accretion flow in quiescent black-hole binaries (Narayan, McClintock, \& Yi 1996).

\section{Conclusions}

The All-Sky Monitor on \it RXTE \rm is exploring and revealing long term variations that were either unknown or poorly known. It has been instrumental in the productive use of the large Proportional Counter Array, and it provides a valuable spectral and temporal context for observations with the PCA and other instruments or telescopes. Several long term periodicities have been found or better established. Some of these may be other examples of accretion disk precession. The ASM light curves for the possibly uniquely young system Cir X-1 and of two superluminal microquasars show complex temporal-spectral behavior either for the first time or in remarkable new detail.

Transitions and other variabilty in PCA data from the very different systems, Cir X-1 and Cyg  X-1, are providing new insight into the conditions near the presumed neutron star or, respectively, presumed black hole.

The PCA has revealed two categories of oscillations near 1 kHz in Low-Mass X-ray Binary (LMXB) systems. Sco X-1 typifies one category, namely a relatively low-Q oscillation (i.e., a QPO) that increases in frequency with mass accretion rate and which is sometimes accompanied by a second QPO at a frequency several hundred Hz lower. The difference between these two frequencies may well be the spin frequency of the neutron star. There are currently eight examples of variable $\sim$1 kHz oscillation.

The other category consists of three sources where pulsations have been observed during X-ray Type-I (thermonuclear) bursts. These are quite coherent and are very likely a direct view of the neutron-star spin. One of these sources (4U 1728--34) exhibits both phenomena: the doublet near 1 kHz and coherent oscillations during a burst. The frequency during the burst agrees well with the difference frequency of the doublet, as expected if both represent the neutron-star spin.

It appears therefore that \it RXTE \rm has indeed demonstrated that the neutron stars in LMXB have been spun up to the high spin rates characteristic of the millisecond radio pulsars. 

The study of microquasars has become a major activity of \it RXTE \rm. In GRS 1915+105, multiple temporal/spectral states, large-amplitude ringing, strong variable QPO, and a recurrent 67-Hz signal have been observed. In GRO 1655--40, a $\sim$298-Hz signal has been found. These latter signals have occurred when the source spectrum is at its hardest. These phenomena in conjunction with observations at other wavelengths should go far toward revealing the role accretion processes play in (radio) jet formation.

In summary, the spectral and temporal phenomena being revealed by \it RXTE \rm are informing us directly about accretion processes, the evolution of compact binary systems, and the nature of the neutron stars and black holes. I close by reminding the reader that there are many other important results from \it RXTE \rm that have not been covered here.

\section*{Acknowledgments}

The author is grateful for the efforts of the entire \it RXTE \rm team and the many observers whose work contributed toward this perspective. This work was supported in part by NASA under contract NAS5-30612. 

\section{References}

\noindent Belloni, T., et al. 1997, ApJ Lett., submitted

\noindent Belloni, T., Mendez, M., van der Klis, M., Hasinger, G., Lewin, W. H. G., \& van Paradijs, J. 1996, ApJ 472, L107.

\noindent Berger, M. et al. 1996, ApJ, 469, L13

\noindent Bhattacharya, D. 1995, in X-ray Binaries, ed. W. H. G. Lewin, J. van Paradijs, \& E. P. J. van den Heuvel (Cambridge: Cambridge Univ. Press), 233

\noindent Bradt, H. V., Rothschild, R. E., \& Swank, J. H. 1993, A\&AS, 97, 355

\noindent Brandt, W. N., Fabian, A., Dotani, T., Nagase, F., Inoue, H., Kotani, T., \& Segawa, Y. 1996, MNRAS 283, 1071

\noindent Castro-Tirado, A. J., Brandt, S., \& Lund, N. 1992, IAU Circ. 5590

\noindent Chen, X, Swank, J. H., \& Taam, R. E. 1997, ApJ Lett., in press

\noindent Corbet, R. 1996, IAU Circ. 6508

\noindent Corbet, R., \& Finley, J. P. 1996, IAU Circ. 6522

\noindent Cui, W., Heindl, B., Rothschild, R. E., Zhang, S. N., Jahoda, K., \& Focke, W. 1997a, ApJ 474, L57

\noindent Cui, W., Zhang, S. N., Focke, W., \& Swank, J. H. 1997b, ApJ, in press

\noindent Dower, R., Bradt, H., \& Morgan E. 1982, ApJ, 261, 228

\noindent Ford, E., et al. 1997, ApJ, 475, L123

\noindent Glass, I. 1994, MNRAS, 268, 742

\noindent Greiner, J., Morgan E. H., \& Remillard, R. A. 1996, ApJ, 473, L107

\noindent Gruber, D. E., \& Rothschild, R. E. 1984, ApJ, 283, 546

\noindent Hjellming, R. M. \& Rupen, M. P. 1995, Nature 375, 464

\noindent Hjellming, R. M., Rupen. M. P., Marti, J., Mirabel, F., \& Rodriguez, L. F. 1996, IAU Circ. 6383

\noindent Klein, R. I., Arons, J., Jernigan, J. G., Hsu, J. J.-L. 1996, ApJ, 457, L85 

\noindent Levine, A. M., Bradt, H., Cui, W., Jernigan, J. G., Morgan, E. H., Remillard, R. Shirey, R. E., \& Smith, D. A. 1996, 469, L33

\noindent McDermott, P. N., Van Horn, H. M., \& Hansen, C. J. 1988, ApJ, 325, 725

\noindent Miller, M. C., Lamb, F. K., \& Psaltis, D. 1997, ApJ, submitted

\noindent Mineshige, S., \& Wheeler, J. C. 1989, ApJ, 343, 241

\noindent Mirabel, I. F., \& Rodriguez, L. F. 1994, Nature 371, 46

\noindent Morgan, E. H., Remillard, R. A., \& Greiner, J. 1997, ApJ, in press

\noindent Narayan, R., McClintock, J. E., \& Yi, I. 1996, ApJ, 457, 821

\noindent Nowak, M. A., Wagoner, R. V., Begelman, M. C., \& Lehr, D. E. 1997, ApJ Lett., in press

\noindent Oosterbroek, T., van der Klis, M., Kuulkers, E., van Paradijs, J., \& Lewin, W. H. G. 1995, A\&A, 297, 141

\noindent Orosz, J. A., \& Bailyn, C. D. 1997, ApJ, in press

\noindent Orosz, J. A., Remillard, R. E., Bailyn, C. D., and McClintock, J. E. 1997, ApJ Lett., in press

\noindent Parmar, A. N., Stella, L., \& White, N. E. 1986, ApJ, 304, 664

\noindent Peele, A. G., \& White, N. E. 1996, IAU Circ. 6524

\noindent Priedhorsky, W. C., \& Holt, S. S. 1987, Space Sci. Rev., 45, 291

\noindent Remillard, R. 1996, Talk presented at the 18th Texas Symposium on Relativistic Astrophysics, Chicago, December 1996; also  Remillard, R., McClintock, J., Bailyn, C., Orosz, J., \& Morgan, E. 1997, ApJ, in preparation

\noindent Shirey, R. E., Bradt, H. V., Levine, A. M., \& Morgan, E. H. 1996, ApJ, 469, L21

\noindent Smale, A. P., \& Lochner, J. 1992, ApJ, 395, 582

\noindent Smale, A. P., Zhang, W., \& White, N. E. 1996, IAU Circ. 6507

\noindent Smarr, L., \& Blandford, R. 1976, ApJ, 207, 574

\noindent Smith, D. A., Morgan, E. H., Bradt, H. 1997, ApJ Lett., in press

\noindent Stewart, R. T., Caswell, J. L., Haynes, R. F., \& Nelson, G. J. 1993, MNRAS, 261, 593

\noindent Strohmayer, T., Lee, U., Jahoda, K. 1996, IAU Circ. 6484

\noindent Strohmayer, T. E., Zhang, W., Swank, J. H., Smale, A., Titarchuk, L., Day, C., \& Lee, U. 1996, ApJ, 469, L9

\noindent Tennant, A. 1988,  MNRAS, 230, 403

\noindent Tingay, S. J., et al. 1995, Nature, 374, 141

\noindent Tsunemi, H., Kitamoto, S., Manabe, M., Miyamoto, S., \& Yamashita, K. 1989, PASJ, 41, 391

\noindent van der Klis, M. et al. 1996b, IAU Circ. 6511

\noindent van der Klis, M., Swank, J., Zhang, W., Jahoda, K., Morgan, E. H., Lewin, W. H. G., Vaughan, B., van Paradijs, J. 1996a, ApJ, 469, L1

\noindent van Paradijs, J. 1995, in X-ray Binaries, ed. W. H. G. Lewin, J. van Paradijs, \& E. P. J. van den Heuvel (Cambridge: Cambridge Univ. Press), 536

\noindent Vargas, M. et al. 1977, ApJ Lett., in press.

\noindent Wijnands et al. 1996b, IAU Circ. 6447

\noindent Wijnands, R. A. D., Kuulkers, E., \& Smale, A. P. 1996a, ApJ 473, L45

\noindent Zhang, S. N., Wilson, C. A., Harmon, B. A., Fishman, G. J., Wilson, R.B., Paciesas, W. S., Scott, M., \& Rubin, B. C. 1994, IAU Circ. 6046

\noindent Zhang, S. N., Cui, W., Harmon, B. A., Paciesas, W. S., Remillard, R. E., van Paradijs, J., \& Yu, W. 1997, ApJ Lett., in press

\noindent Zhang, S. N., Robinson, C. R., \& Cui, W. 1996a, IAU Circ. 6510

\noindent Zhang, W., Lapidus, I., White, N. E., \& Titarchuk, L. 1996, ApJ, 469, L17

\section {Figure Captions}

\noindent Figure 1: Six light curves (1.5--12 keV) from the ASM for the year 1996. In each plot, the ordinate is ASM ct/s; 1.0 Crab = 75 ASM ct/s. From A. Levine, pvt. communication.

\vspace{0.5cm}
\noindent Figure 2: Six cycles of the 16.6-d period of Cir X-1 with hardness ratios from the ASM (1.5--12 keV). 1.0 Crab = 75 ASM ct/s. From Shirey et al. 1996, ApJ, 469, L21.

\vspace{0.5cm}
\noindent Figure 3: Segments of PCA data from Cir X-1 with 1-s time bins at three different phases of the 16-d periodicity. Note the dramatic dips in (a) and (b) and the fluctuations which greatly exceed counting statistics in (c). From Shirey et al. 1996, ApJ, 469, L21.

\vspace{0.5cm}
\noindent Figure 4: Power density spectra from Cir X-1 (PCA data) during the quiescent phases of one 16.6-d cycle. Each PDS is plotted one decade below the previous one. They are ordered by QPO frequency which approximates orbital phase. From Shirey et al. 1996, ApJ, 469, L21.

\vspace{0.5cm}
\noindent Figure 5: Effective time lags of hard photons from Cyg X-1 for three bands with effective energies, 3, 9, and 33 keV respectively from PCA data. The lags of the hard band (asterisks) and medium band (diamonds) relative to the soft band are shown. The lags are the average over 1--10 Hz in PDS frequency. From Cui et al. 1997, ApJ, in press.

\vspace{0.5cm}
\noindent Figure 6: Power density spectrum for Sco X-1 from the PCA showing the two peaks (upper) in one data set, and the single peak (lower). From van der Klis et al. 1996a, ApJ, 469, L1.

\vspace{0.5cm}
\noindent Figure 7: Frequency of QPO from 4U 1728--34 during the burst shown in the lower panel (PCA data). Note the stability of the frequency during the tail at 364 Hz. From T. Strohmayer, pvt. communication.

\vspace{0.5cm}
\noindent Figure 8: Power density spectrum (PCA data) for a 2-s period during a Type-I X-ray burst from KS 1731--260. From Smith, Morgan, \& Bradt 1997, ApJLett, in press.

\vspace{0.5cm}
\noindent Figure 9: Expanded ASM light curve (1.5--12 keV) of microquasar GRS 1915+105 showing observation times of PCA (tic marks above), the several different states with their names, and hardness ratios (defined in Fig. 2). 1.0 Crab = 75 ASM ct/s. From Morgan, Remillard, and Greiner 1996, ApJ, in press.

\vspace{0.5cm}
\noindent Figure 10: PCA light curve from microquasar GRS 1915+105 showing dramatic dips and flaring. Note the narrow precursor dips in the upper figure and the softening of the spectrum during the dips. This and the excess flux after some dips argue against absorption as the cause of this variability. From Greiner, Morgan and Remillard 1996, ApJ, 473, L107.

\vspace{0.5cm}
\noindent Figure 11: Power density spectrum from GRS 1915+105 for 1996 May 5 (PCA data). The pronounced peak at 0.067 Hz (15 s) has 3 harmonics. Note also the peak at 67 Hz. From E. Morgan, pvt. communication.

\vspace{0.5cm}
\noindent Figure 12: Power density spectra for GRS 1915+105 showing the six appearances of the 67-Hz feature. From Morgan, Remillard, and Greiner 1996, ApJ, in press.

\vspace{0.5cm}
\noindent Figure 13: ASM light curve (1.5--12 keV) of the microquasar GRO 1655--40. 1.0 Crab = 75 ASM ct/s. From D. A. Smith, pvt. communication.

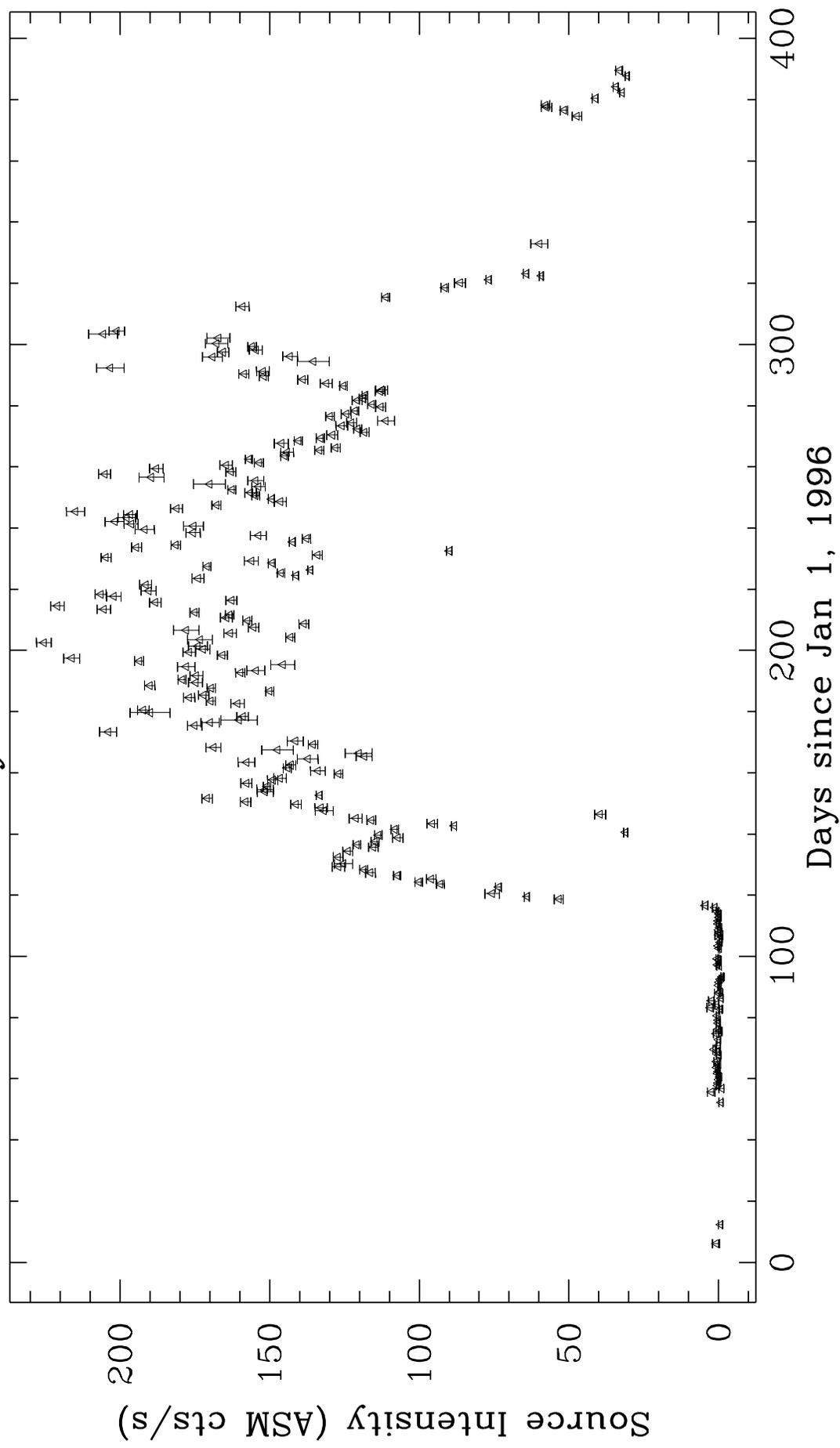
\end{document}